\documentclass[aps,prb,reprint,twocolumn,showpacs,floatfix,superscriptaddress]{revtex4-2}

\usepackage{amssymb,amsmath,amstext}                
\usepackage{graphicx}                                               
\usepackage{epstopdf}                                               
\usepackage{color}                                                     
\usepackage{bm}                                                        
\usepackage{appendix}                                              
\usepackage[utf8]{inputenc}
\usepackage{latexsym}
\usepackage{xcolor}
\usepackage{braket}
\usepackage{ulem}
\usepackage{siunitx}
\usepackage{umoline}
\usepackage{epsfig}
\usepackage{graphics}
\usepackage{amssymb}
\usepackage{psfrag}
\usepackage{mathtools}
\usepackage[english]{babel}
\usepackage[T1]{fontenc}
\usepackage{lmodern}
\usepackage{booktabs}

\usepackage[colorlinks, linkcolor=blue, anchorcolor=blue, citecolor=blue]{hyperref}

\begin{document}

\title{The role of interaction-induced tunneling in the dynamics of polar lattice bosons}%
\author{Adith Sai Aramthottil} 
\affiliation{Institute of Theoretical Physics, Jagiellonian University in Krakow, ul. Lojasiewicza 11,
	30-348 Krak\'ow, Poland}
\author{Mateusz \L\c{a}cki} 
\affiliation{Institute of Theoretical Physics, Jagiellonian University in Krakow, ul. Lojasiewicza 11,
	30-348 Krak\'ow, Poland}

\author{Luis Santos}
\affiliation{Institut f\"ur Theoretische Physik, Leibniz Universit\"at Hannover, Germany}
\author{Jakub Zakrzewski}
\email{jakub.zakrzewski@uj.edu.pl}
\affiliation{Institute of Theoretical Physics, Jagiellonian University in Krakow, ul. Lojasiewicza 11,
	30-348 Krak\'ow, Poland}
\affiliation{Mark Kac Complex Systems Research Center,  Jagiellonian University in Krakow, \L{}ojasiewicza 11, 30-348 Krak\'ow, Poland}

\date{\today}

\begin{abstract}
Inter-site dipolar interactions induce, even in absence of disorder, an intriguing non-ergodic dynamics for dipolar bosons in an optical lattice. We show that the inherent dipole-induced density-dependent tunneling, typically neglected, plays a crucial role in this dynamics. For shallow-enough lattices, the delocalization stemming from the interaction-induced hopping overcomes the localization induced 
by inter-site interactions. As a result, in stark contrast to the more studied case of hard-core bosons, delocalization is counter-intuitively strengthen when the dipolar strength increases. Furthermore, the quasi-cancellation between bare and interaction-induced tunneling may lead, near a lattice-depth-dependent value of the dipole strength, 
to an exact decoupling of the Hilbert space between ergodic hard-core states and strongly non-ergodic soft-core ones. Our results show that interaction-induced hopping should play a crucial role in future experiments on the dynamics of polar lattice gases.
\end{abstract}

\maketitle

\section{Introduction}

Many-body localization~(MBL) has attracted in recent years a major attention as a paradigmatic manifestation of nonergodic dynamics in the presence of disorder~\cite{Alet18,Abanin19,Gopalakrishnan20}. 
While the very existence of MBL in the thermodynamic limit remains a controversial and extensively discussed topic~\cite{Suntajs20e,Panda20,Abanin21,Sierant20,Sierant20p,Suntajs20,Sels21,Sierant21,Crowley22,Morningstar22,Sierant22}, 
experimental signatures of nonergodic dynamics in finite systems on a time scale of several hundreds of tunneling times have been clearly observed~\cite{Schreiber15,Luschen17,Roushan17,Xu18,Rispoli19,Lukin19}.
Recent years have brought also a number of examples of nonergodic dynamics in disorder-free systems, ranging from implementations of lattice gauge theories~\cite{Brenes18,James19,Robinson19}, to tilted lattices and smooth potentials~\cite{Scherg21,Taylor20,Guo21,Morong21,Yao20,Yao21}. A prominent example, related to an approximate global constraint and an appropriate choice of the initial state, is given by the so-called 
quantum scars~\cite{Turner18,James19,Szoldra22,Adith22}. Approximate global constraints result often in Hilbert-space fragmentation \cite{Khemani20}.

A particularly interesting example of Hilbert-space fragmentation and disorder-free nonergodic dynamics is provided by polar gases in optical lattices~\cite{Li21}.  A sufficiently-large dipole strength results in an emerging dynamical constraint given by the approximate conservation of {the number of pairs of nearest-neighbor~(NN) particles}. This, combined with the eventual conservation of the number of next-to-NN pairs, results in Hilbert space shattering~\cite{Li21} and strongly nonergodic dynamics in hard-core systems. In those systems, on-site interactions are assumed large-enough to prevent more than one particle per lattice site.
 
In this work we show that the dynamics of soft-core dipolar bosons, with possibly multiply-occupied sites, may be drastically different than their hard-core counterparts. 
This marked difference results from the crucial role played by interaction-induced density-dependent tunneling~(DDT).  
Although DDT may be generally relevant in Hubbard models with strong-enough on-site interactions~\cite{Hirsch94, Dutta15}, it is particularly relevant
in polar lattice gases due to the long-range dipole-dipole interactions, as shown by recent studies of their ground-state properties~\cite{Maik13,Biedron18,Kraus20,Kraus22}. 
Our results show that due to DDT, a growing dipole strength results in enhanced particle delocalization, in a stark contrast to the hard-core case. Moreover,  DDT induces, for a particular, lattice-depth-dependent dipolar strength, a quasi-cancellation between kinetic tunneling and DDT leading to a peculiar exact decoupling of the Hilbert space into ergodic and strongly non-ergodic states.

The structure of the paper is as follows. Section~\ref{sec:Model} introduces the lattice model under consideration. Section~\ref{sec:Spectral} is devoted to the spectral properties of soft-core bosons. Section~\ref{sec:Dynamics} discusses how the effect of DDT on the spectral properties translates into a markedly modified particle dynamics. In Sec.~\ref{sec:Critical} we study the case in which the bare hopping and the DDT quasi-cancel. Finally, in Sec.~\ref{sec:Conclusions}, we summarize our conclusions.


\section{Model} 
\label{sec:Model}
We consider externally-oriented dipolar bosons in a deep one-dimensional optical lattice. The system is well described by the extended Bose-Hubbard~(EBH) model:
\begin{eqnarray}
\hat{\mathcal{H}}_{EBH} &=& -t\sum_{j=1}^{L-1} \left (\hat{a}^{\dagger}_j\hat{a}_{j+1}+\mathrm{H.c.} \right )+\frac{U}{2}\sum_{j=1}^L\hat{n}_j(\hat{n}_j-1)\nonumber \\
&+&\frac{V}{2}\sum_{i\neq j}\frac{1}{\vert i-j \vert^3  }\hat{n}_i\hat{n}_j  \nonumber \\
&-&T\sum_{j=1}^{L-1} \left [\hat{a}^{\dagger}_j(\hat{n}_j+\hat{n}_{j+1})\hat{a}_{j+1}+\mathrm{H.c.} \right ].
\label{eq:ext}
\end{eqnarray}
where $a^\dagger_j$~($a_j$) denotes the bosonic creation~(annihilation) operator, and $t$ is the bare tunneling amplitude. The first line of Eq.~\eqref{eq:ext} 
is the standard Bose-Hubbard model, in which the on-site interaction strength $U$ results from both contact-like and dipole-dipole interactions. We fix $U/t=3$ below. 
The second line describes the inter-site dipolar interactions characterized by the dipolar strength $V$, which may be tuned by changing the dipole orientation with respect to the lattice axis. 
We note in passing, that keeping $U$ fixed while changing $V$ requires tuning the contact interaction, via e.g. a Feshbach resonance. We assume a strong confinement transversal to the lattice axis, since otherwise the $1/r^3$ decay of the dipolar interaction should be generally modified~\cite{Wall13,Korbmacher23}.

\begin{figure}[t!]
	\includegraphics[width=0.7\linewidth]{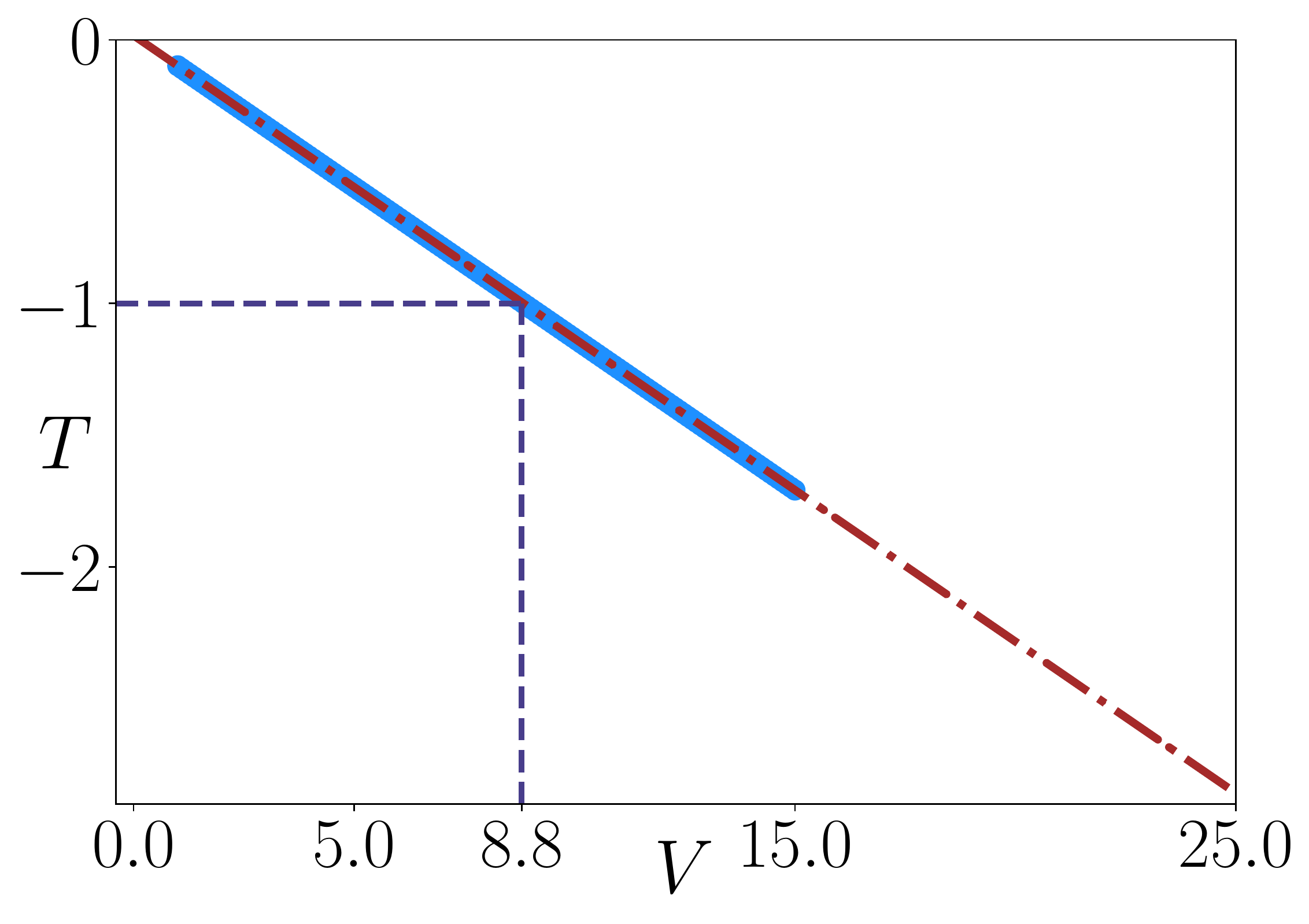}
	\caption{Relation between the DDT rate $T$ and $V$ for $U/t=3$ and $s=8$. The dash-dotted line shows the almost perfect correspondence with a linear function. The dashed lines emphasize that $T/t=-1$ at $V/t=8.8$.}
	\label{fig:1}
\end{figure}

The last line in Eq.~\eqref{eq:ext} corresponds to the DDT, with amplitude $T$, which, interestingly, is negative. For a given lattice depth $V_0$~(which we characterize below by $s=V_0/E_{\rm R}$, with $E_{\rm R}$ the recoil energy), and employing the appropriate form of the on-site functions~(see App.~\ref{sec:AppA}), one finds that, for the moderate value of $U/t$ considered, $T$ is linearly proportional to $V$ over a broad range of $V/t$ values~(Fig.~\ref{fig:1}). Particularly relevant, as discussed below, is the case in which 
$T/t=-1$, which occurs for an $s$-dependent critical $V/t$.



We consider in the following a half-filled lattice, with $N=L/2$ bosons in $L$ sites, with open boundary conditions. This choice facilitates the comparison with previous studies on hard-core bosons~\cite{Li21}. Since the maximal site occupation equals the total particle number, we are limited in our exact diagonalization analysis to system sizes up to $L=16$~(which corresponds to a large Hilbert space dimension of 490314 states). Although this precludes a reliable extrapolation to the thermodynamic limit, it provides already a clear qualitative picture, and it is quantitatively relevant for site-resolved experiments on ultra-cold gases in optical lattices, also typically limited to a small number of sites~\cite{Lukin19,Rispoli19}.


\section{Spectral properties}
\label{sec:Spectral}

In this section, we discuss how DDT radically modifies the spectral properties of soft-core polar lattice gases. 


\subsection{Density of states}

\begin{figure}[t!]
\includegraphics[width=0.8\linewidth]{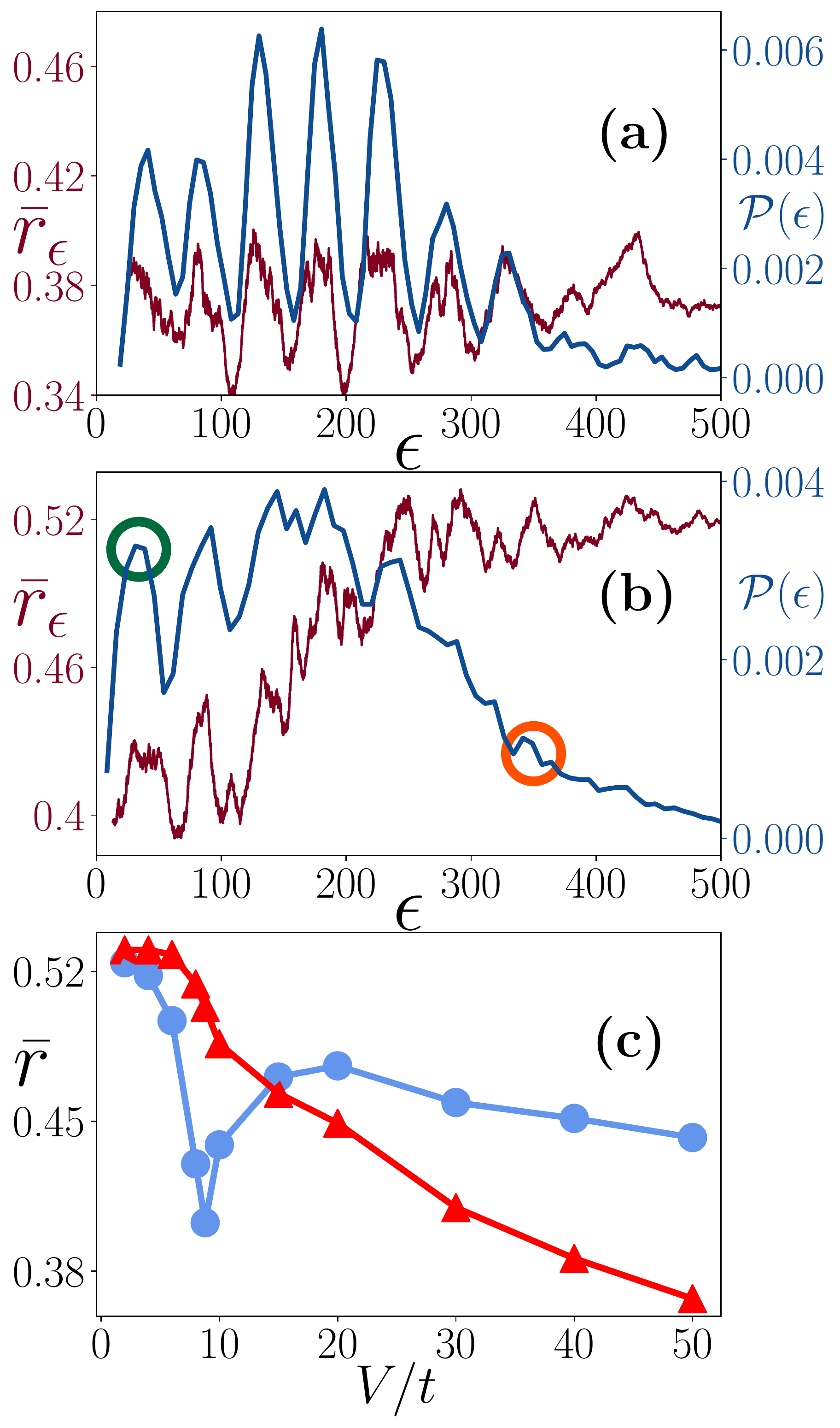}
\caption{Energy dependence of the DoS, ${\cal P}(\epsilon)$, and the gap ratio, $\overline{r}_\epsilon$, for $T=0$~(a) and in the presence of DDT~(b) for $L=16$ with $V/t=50$. The circles indicate the initial states employed in Fig.~\ref{fig:4}(b). (c) The mean gap ratio, $\overline{r}$, for $s=8$ and $L=14$ as a function of $V/t$ with~(circles) and without~(triangles) DDT.}	
\label{fig:2}
\end{figure}

Hard-core dipolar bosons undergo Hilbert-space shattering for large-enough $V/t$ due to the emergent constraint induced by the approximate conservation of the number of NN and next-to-NN pairs~\cite{Li21}. A similar behavior is shared by soft-core bosons in the absence of DDT. Figure~\ref{fig:2}(a) 
shows the density of states~(DoS), ${\cal P}(\epsilon)$, 
for $L=16$ with $V/t=50$ and $s=8$. The DoS presents pronounced peaks corresponding to different number of occupied NN links, $N_{NN}=\sum_{j=1}^{L-1} \langle n_j n_{j+1} \rangle$. In contrast, in the presence of DDT, Hilbert-space fragmentation is largely washed out~(Fig.~\ref{fig:2}(b)), even for large $V/t$, indicating the lack of conservation of $N_{NN}$. 



\subsection{Mean gap ratio} 

The DDT also strongly modifies the level-spacing statistics. This is best observed in the behavior of the gap ratio, defined as $r_n={\mathrm {min}}(\delta_n, \delta_{n+1})$ where $\delta_n=E_n-E_{n-1}$ with $\{E_n\}$ being the ordered set of eigen-energies~\cite{Oganesyan07}. The mean gap ratio $\overline{r}$ is evaluated as the average value over the whole spectrum. In 
Fig.~\ref{fig:2}(c), 
we depict $\overline{r}$ for $L=14$, $s=8$ and different ratios $V/t$. 
Integrable systems, with Poissonian level statistics, are characterized by $\overline{r}\approx 0.383$, whereas for ergodic time-reversal invariant systems one expects $\overline{r}\approx 0.53$, corresponding to the Gaussian Orthogonal Ensemble~(GOE) of random matrices. We observe the latter behavior only for low $V/t$. For increasing $V/t$, a general decrease of $\overline{r}$ is observed. Whereas for $T=0$ the gap ratio reaches a Poissonian value for large $V/t$, signaling quasi-integrability, the presence of DDT results in a mixed dynamics, with the value of $\overline{r}$ lying in between Poissonian and GOE statistics. Note as well the pronounced sharp minimum at $V/t=8.8$, which is related to the condition $T/t=-1$, discussed in detail in Sec.~\ref{sec:Critical}.



\begin{figure}[t!]
\includegraphics[width=\linewidth]{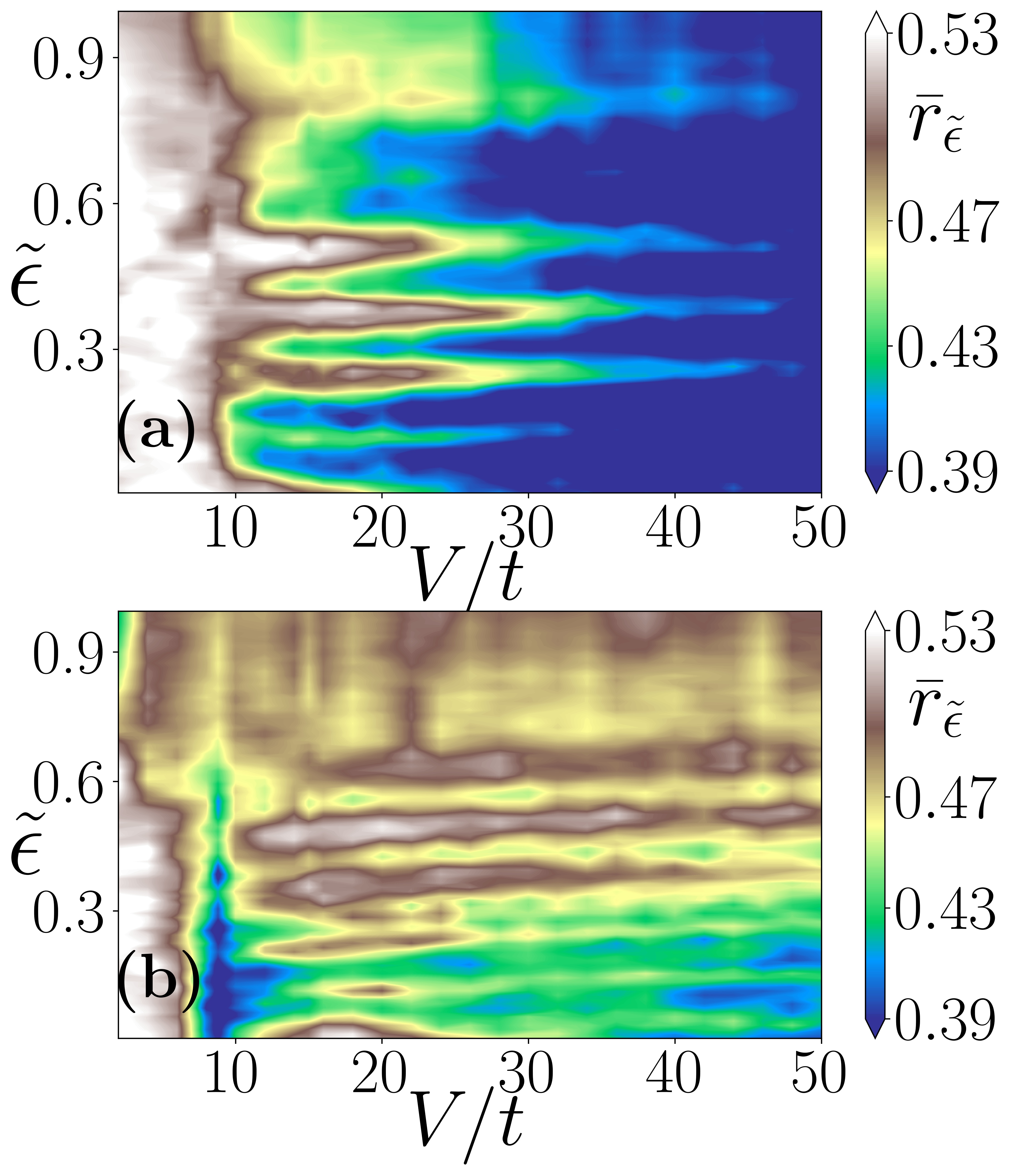}
\caption{Energy dependent gap ratio, $\bar r_\epsilon$, as a function of the dipolar interaction strength, $V/t$, and the scaled energy, $\tilde \epsilon$, for $L=14$ sites and $s=8$ with $T=0$~(a) and with DDT~(b).}	
\label{fig:3}
\end{figure} 



\subsection{Energy dependence of the gap ratio}
The $r_n$ values present a marked energy dependence, which plays a crucial role in the dynamics~(see Sec.~\ref{sec:Dynamics}).
In order to monitor this dependence, we introduce the scaled energy $\tilde\epsilon\in[0,1]$, defined as $\tilde\epsilon = (\epsilon-\epsilon_{\textrm{min}})/(\epsilon_{\textrm{max}}-\epsilon_{\textrm{min}})$, where $\epsilon_{\textrm{min}}$ ($\epsilon_{\textrm{max}}$) is the minimum (maximum) 
eigenenergy for a given $V/t$. 
We determine the gap ratio $\bar r_\epsilon$ as the rolling average of $4000$ energy gaps in the scaled energy interval around a given $\tilde \epsilon$. 

In the absence of DDT~(see Fig.~\ref{fig:3}(a)), the whole spectrum is ergodic at low $V/t$. At $V/t\approx 10$, a strong energy dependence appears in the form of approximately ergodic fingers separated by approximately regular regions. This structure correlates with the modulation of the density of states~(see Fig.~\ref{fig:2}(a)).  

The situation is markedly different in the presence of DDT~(see Fig.~\ref{fig:2}(b) and Fig.~\ref{fig:3}(b)). High-lying eigenstates remain ergodic even for very large $V/t$ ratios. Only low-lying states are significantly nonergodic, with a gap ratio close to the Poissonian value, since these states are characterized by a small number of NN links, reducing the effective role played by the DDT. At the largest $V/t$ values, approximately one half of the states belongs to the ergodic sector, explaining the fact that $\overline{r}$ saturates around $0.45$~(Fig.~\ref{fig:2}(c)). This strongly suggests that the dynamics of polar lattice gases initially built from low-energy eigenstates must be markedly different from that of systems prepared in high-energy ones. Whereas the former should reveal localization features, the latter should present ergodic dynamics. 
Note as well, that the spectrum becomes to a large extent Poissonian at $V/t=8.8$~(for $s=8$), corresponding with the dip in $\overline{r}$ observed in Fig.~\ref{fig:2}(c).



\begin{figure}%
	\includegraphics[width=\linewidth]{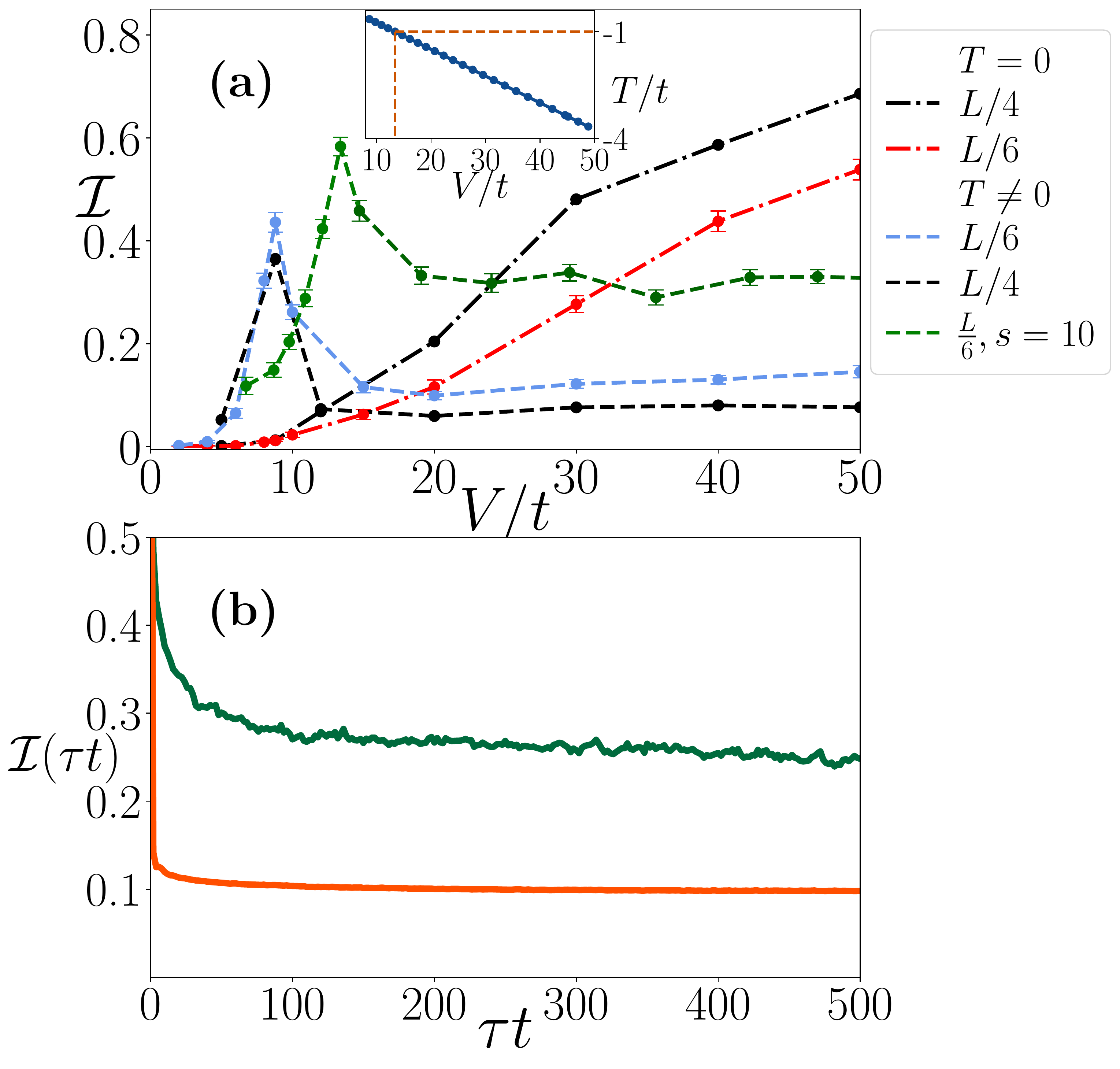}  
	\caption{(a) Inhomogenity $\mathcal{I}(\tau_f=500/t)$ as a function of $V/t$, for $L=12$ and $s=8$. The curves~(see legend) depict different cases without~($T=0$) and with DDT~($T\neq 0$), for initial states with $N_{NN}=L/6$ and $L/4$. We depict as well the case with $s=10$ and $N_{NN}=L/6$. Observe that the peak of enhanced inhomogeneity is at $V/t=8.8$ for $s=8$ but at $V/t\approx 13$ for $s=10$, corresponding to the different value of $V/t$ at which $T/t=-1$~(see inset for the dependence of $T/t$ on $V/t$ for $s=10$). Error bars indicate the result from the bootstrap estimate based on data for about $200$ initial conditions in each sector. (b) Inhomogeneity ${\cal I}(\tau)$ for $L=16$ with $V/t=50$, for initial states with two different energies corresponding to the cases indicated by circles in Fig.~\ref{fig:2}(b).}
	\label{fig:4}
\end{figure} 



\section{Particle dynamics}
\label{sec:Dynamics}

In this section, we show how the modified spectral properties translate into a radically altered dynamics in the presence of DDT. 


\subsection{Homogeneity}

We consider the time evolution of the system when starting with an initial Fock state, $|\varphi\rangle$. In view of the expected energy-dependence of the dynamics, we choose carefully the initial state such that its energy lies at the center of an energy window with a large DoS, 
avoiding regions of low density of states occurring due to a possible Hilbert-space fragmentation. Whereas recent MBL experiments have 
employed an initial density wave~\cite{Schreiber15,Luschen17}, 
this is not a good choice for a polar lattice gas with large $V/t$, 
since it lies at the extremes of the spectrum. 
Instead, we consider a manifold of initial Fock states, with a given number of NN pairs, $N_{NN}$.

We are interested in how the initial inhomogeneous population in the lattice redistributes at time $\tau>0$ amongst the sites, and in particular whether it becomes eventually homogeneous, washing out any information about the initial distribution. Density homogenization is best analyzed using the inhomogeneity parameter:
\begin{eqnarray}
\mathcal{I}(\tau) = \frac{\sum_{i=1}^L\left ( \langle \hat{n}_i(\tau) \rangle - \rho \right )^2}{\sum_{i=1}^L\left( \langle \hat{n}_i(0) \rangle -\rho \right )^2},
\label{eq:zak_fac}
\end{eqnarray}
with $\rho=N/L$~($=1/2$ in our case). Note that $0<\mathcal{I}<1$, with $0$~($1$) indicating a fully homogeneous~(inhomogeneous) distribution.

Figure~\ref{fig:4} shows $\mathcal{I}(\tau_f)$ for $L=12$ and different $V/t$ ratios, after an experimentally accessible time $\tau_f=500/t$~\cite{Scherg21}. We obtain $\mathcal{I}(\tau_f)$ after averaging over initial Fock states with $N_{NN}=L/4$, which is the  most populated sector in the possibly fragmented Hilbert space~\cite{Li21}. In Fig.~\ref{fig:4}, we depict as well our results for $N_{NN}=L/6$ for comparative purposes.  As for the case of hard-core bosons, in the absence of DDT, when $V/t$ grows the 
inhomogeneity at a fixed time increases, indicating the strongly non-ergodic character of the dynamics. In contrast, the presence of DDT results even for large $V/t$ in a low~(but non zero) inhomogeneity, reflecting the delocalizing role played by the DDT, in agreement with the markedly different spectral properties.

The above-mentioned energy dependence of the level statistics is reflected in the different dynamics observed for initial conditions belonging to different spectral regions. We illustrate this point in Fig.~\ref{fig:4}(b), 
where we depict the evolution of the inhomogeneity 
for $L=16$ with $V/t=50$, and two different initial conditions indicated in Fig.~\ref{fig:2}(b). 
For initial Fock states with $N_{NN}=0$~(corresponding to low energies) ${\cal I}(t)$ remains very 
significant even at $\tau_f=500$, the largest time considered. 
This is in agreement with the fact that the corresponding eigenstates present an approximately Poissonian level statistics~(Fig.~\ref{fig:3}(b)). 
In contrast, the inhomogeneity reaches much lower values for initial states with $N_{NN}=L/4$, which lie at high energies, and are characterized by an approximately GOE level spacing~(Fig.~\ref{fig:3}(b)). Whether the nonzero saturation value of $\cal{I}$ is due to the small system size~(resembling the behavior observed in the disordered XXZ model~\cite{Sierant22}) or to a not fully chaotic behavior cannot be determined with the system sizes studied here. 


\subsection{Dependence on the lattice depth}

In the absence of DDT, the static and dynamic properties of the EBH model 
are given by the value of $V/t$ and $U/t$, irrespective of the actual lattice depth $s$~(which is just relevant for fixing the overall time scale $1/t$). The situation changes when considering the effect of the DDT. The value of $T$ is an $s$-dependent function of $V/t$. For a fixed $V/t$ ratio, $T/t$ decreases when $s$ increases, and hence the effect of DDT is reduced. The dynamics is hence markedly dependent on the lattice depth. 

\begin{figure}%
	\includegraphics[width=0.8\linewidth]{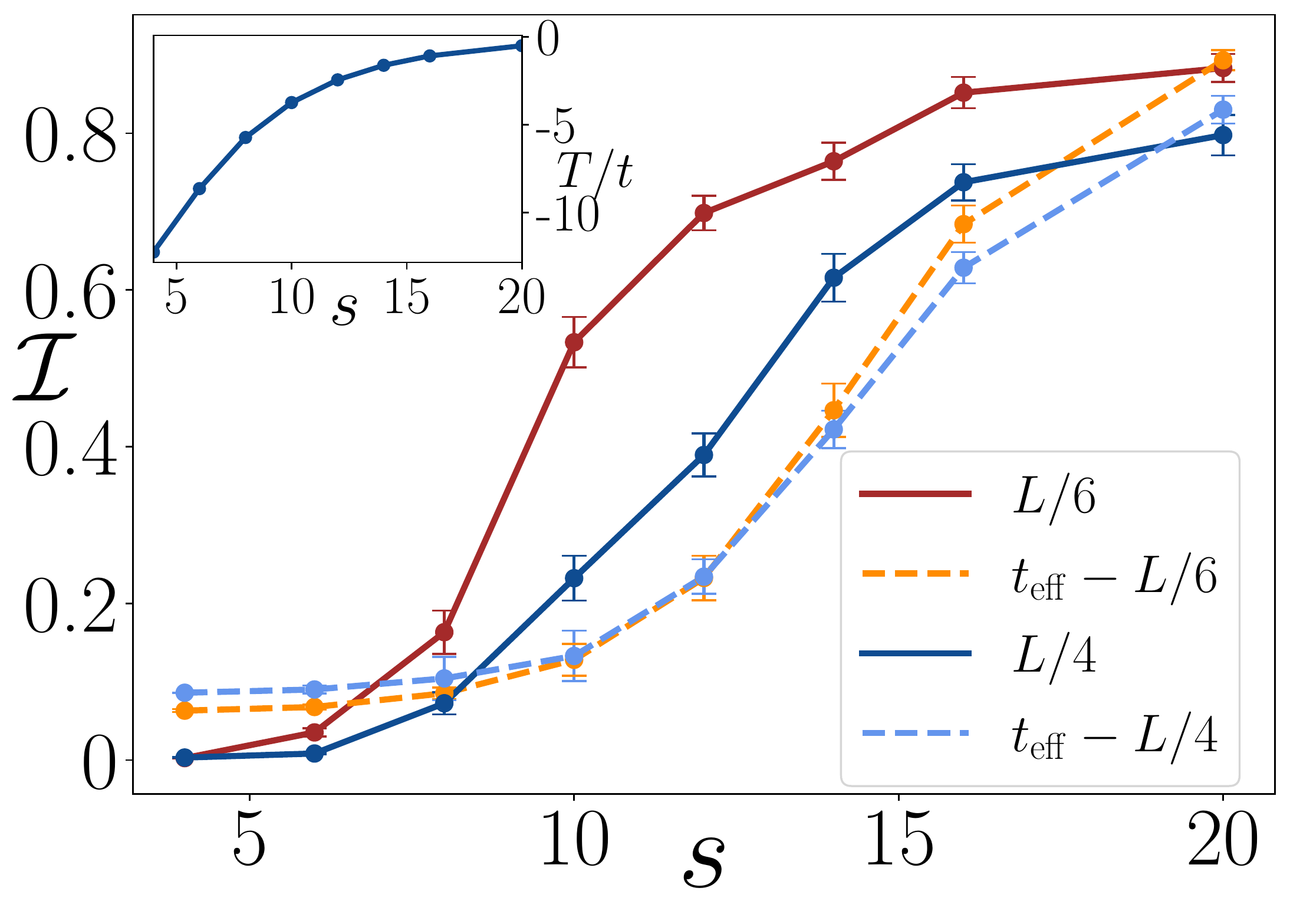}  
	\caption{ Inhomogenity $\mathcal{I}(\tau_f=500/t)$ as a function of the lattice depth, $s$, for $L=12$ with $V/t=50$, for initial states with $N_{NN}=L/4$ (blue solid curve) and $N_{NN}=L/6$ (red solid curve). The dashed lines show the corresponding results in the mean field approximation when DDT is replaced by an effective hopping rate $t_{eff}$. The inset shows $T/t$ as a function of $s$ for the case under consideration.}
	\label{fig:5}
\end{figure}

This dependence is illustrated in 
Fig.~\ref{fig:5}, where we plot as a function of $s$ 
the inhomogeneity ${\cal I}(\tau_f=500/t)$ for $L=12$ with $V/t=50$, and different initial sectors. For low-enough $s$, the DDT is relevant, and the system reaches homogeneity despite the large $V/t$ value. In contrast, for deeper lattices, the delocalizing effect of the DDT becomes less relevant compared to the localizing role of inter-site interactions. As a result, ${\cal I}(\tau_f)$ reaches large values indicating strongly non-ergodic dynamics. 


 
\subsection{Mean-field analysis} 

A qualitative understanding of the effect of the DDT is provided by 
applying mean-field decoupling~\cite{Kraus20}:
\begin{equation}
-\hat a_i^\dag \left [ t + T \left ( \hat n_i + \hat n_j \right ) \right ] \hat a_j \simeq -\left ( t + 2\rho T\right ) \hat a_i^\dag \hat a_j.    
\end{equation}
Since in our case $\rho=1/2$, DDT results in an effective mean-field hopping rate $t_{\mathrm{eff}}=t+T$. It becomes evident that $T/t=-1$ is a special case, which we discuss in Sec.~\ref{sec:Critical}. 
The dynamics 
is hence not regulated by the ratio $V/t$, as in the absence of DDT, but rather by $V/t_{\mathrm{eff}}=\frac{V/t}{1+T/t}$. 
Since $T(V)=\alpha(s)+\beta(s) V$, with $\beta(s)<0$, the ratio approaches $V/t_{\mathrm{eff}}\simeq -\frac{1}{\beta(s)}$ for 
a sufficiently large $V/t$. Hence, increasing the dipolar strength, does not result~(as in the absence of DDT) in a diverging ratio between inter-site interactions and hopping, which leads necessarily to localization, but rather in a saturated ratio,  $|V/t_{\mathrm{eff}}|_{max}$. This maximal ratio depends on the lattice depth, increasing with growing $s$. 
This explains two relevant qualitative features in Fig.~\ref{fig:4} and Fig.~\ref{fig:5}: the independence of $\mathcal{I}$ of $V/t$ for large-enough $V/t$, and the very low inhomogeneity observed even for low $s$. The latter results from the low value of  $|V/t_{\mathrm{eff}}|_{max}$. 

In Fig.~\ref{fig:5} we compare our results using the full EBH in Eq.~\eqref{eq:ext} with those obtained in the mean-field-inspired model in which the DDT is replaced by modifying the kinetic tunneling $t$ into $t_{\mathrm{eff}}$. As expected, the effective mean-field model reproduces well the qualitative features, although there are marked quantitative differences due to the significant density fluctuations in the system.



\begin{figure}%
	\includegraphics[width=0.8\linewidth]{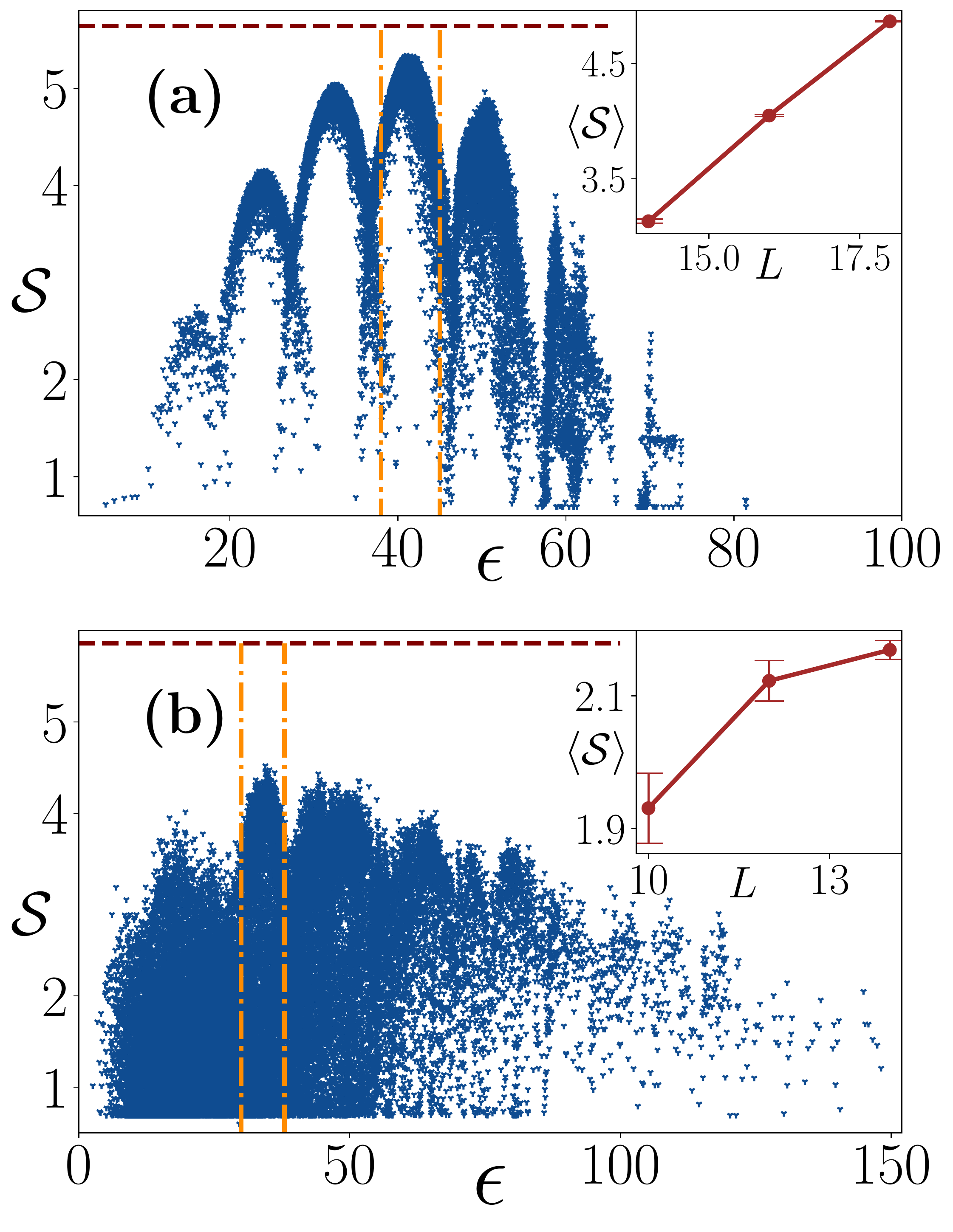}
	\caption{ (a) Entanglement entropy $\mathcal{S}$ for the hard-core sector at $T/t=-1$ ($V/t=8.8$ for $s=8$) for $L=18$. The dashed horizontal line corresponds to the random-matrix theory value. Vertical lines indicate a single sector taken to calculate the average entropy~(shown in the inset for different system sizes). Panel (b) shows  the entanglement entropy for the remaining decoupled sector for  $L=14$. The average entropy does not depend significantly on the system size for this nonergodic sector.}
	\label{fig:6}
\end{figure}



\section{Critical dipole strength}
\label{sec:Critical}
Interestingly, in contrast to what happens for large $V/t$, the presence of DDT may result in a strongly non-ergodic dynamics for relatively modest values of $V/t$, for which the model without DDT would predict ergodicity. As already hinted in previous sections 
this occurs when $T/t\simeq -1$, for which the bare hopping and the DDT quasi-cancel each other~\cite{Kraus22}. This is reflected in the marked maximum observed in the average gap ratio $\overline{r}$ in  Fig.~\ref{fig:2}, and in the inhomogeneity in Fig.~\ref{fig:4}.

For $T/t=-1$, the hard-core Hilbert subspace, with maximally one particle per site, exactly decouples from those states with at least one site with double or higher occupation. 
These two subspaces present markedly different spectral properties. Whereas the hard-core boson subspace presents GOE-like statistics~($\overline{r}\approx0.527$), matching the behavior observed in Ref.~\cite{Li21}, the rest of the Hilbert space~(the soft-core subspace) shows an approximately Poissonian statistics~($\overline{r}\approx 0.4$) as the states in this subspace are strongly affected by the destructive interplay between kinetic and interaction-induced tunneling occuring for $T/t=-1$. This behavior is, as other properties, energy dependent.


\subsection{Eigenstate properties}

\begin{figure}%
	\includegraphics[width=0.9\linewidth]{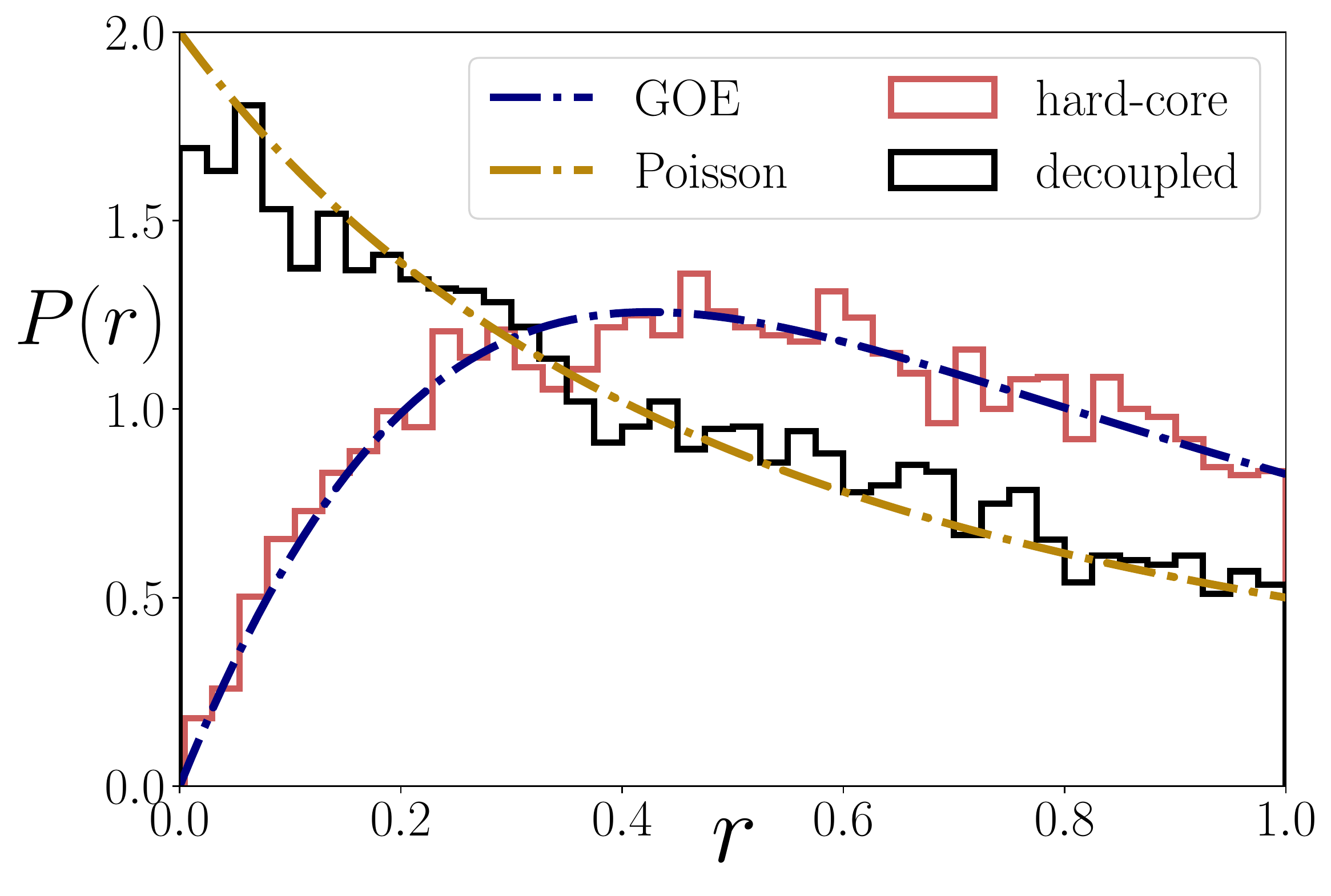}  
	\caption{Histogram $P(r)$ of the gap ratio, $r$, for 
 $s=8$ and $V/t=8.8$~(corresponding to $T/t=-1$). The dashed-dotted lines correspond to the GOE and Poisson predictions. The histograms are presented separately for the two decoupled sectors of hard-core bosons (with GOE statistics) and the remaining soft-core subspace (showing close to Poissonian behavior). The values for $r$ are chosen from within the orange dashed-dotted lines indicated in Fig.~\ref{fig:6}.   In the hard-core sector we consider $L=18$, whereas for the soft-core one~(richer in eigenstates) we are restricted to $L=14$.}
	\label{fig:7}
\end{figure}

Let us consider first the eigenstate properties in both the hard- and soft-core sectors for $T/t=-1$. We focus in particular on the half-chain entanglement entropy, $\mathcal{S}=-Tr[\rho_{L/2}\ln{\rho_{L/2}}]$, where the reduced density matrix $\rho_{L/2} = Tr_{1,\cdots,L/2} \vert \psi \rangle \langle \psi \vert $ 
is obtained after tracing out half of the system for a given eigenstate $ \vert \psi \rangle $.  
Figure~\ref{fig:6}(a) 
shows $\mathcal{S}$ for the hard-core sector for $L=18$ with $V/t=8.8$ and $s=8$, corresponding with $T/t=-1$. It displays a 
finger-like structure due to partial Hilbert-space fragmentation. 
Thus, for the evaluation of the average entropy we consider only eigenstates corresponding to a single finger, as denoted by vertical dotted-dashed lines in Fig.~\ref{fig:6}(a). The hard-core sector is characterized by large entropies, with the average growing linearly with the system size, indicating volume-law scaling, characteristic of delocalized states and nonintegrable dynamics. A similar analysis for the soft-core sector reveals a much broader distribution of $\mathcal{S}$ with many low entanglement states, and a 
a much weaker entropy growth with the system size~(Fig.~\ref{fig:6}(b)). 
This characterisation of the eigenstates nicely matches with the
 gap ratio histograms shown in Fig.~\ref{fig:7}.  The dashed-dotted theoretical predictions drawn in Fig.~\ref{fig:7} are $P(r)=2/(1+r)^2$ for the Poissonian case and $P(r)= 27(r+r^2)/8(1+r+r^2)^{5/2}$, a good approximation for the GOE case~\cite{Atas13}.


 
\begin{figure}	
\includegraphics[width=0.9\linewidth]{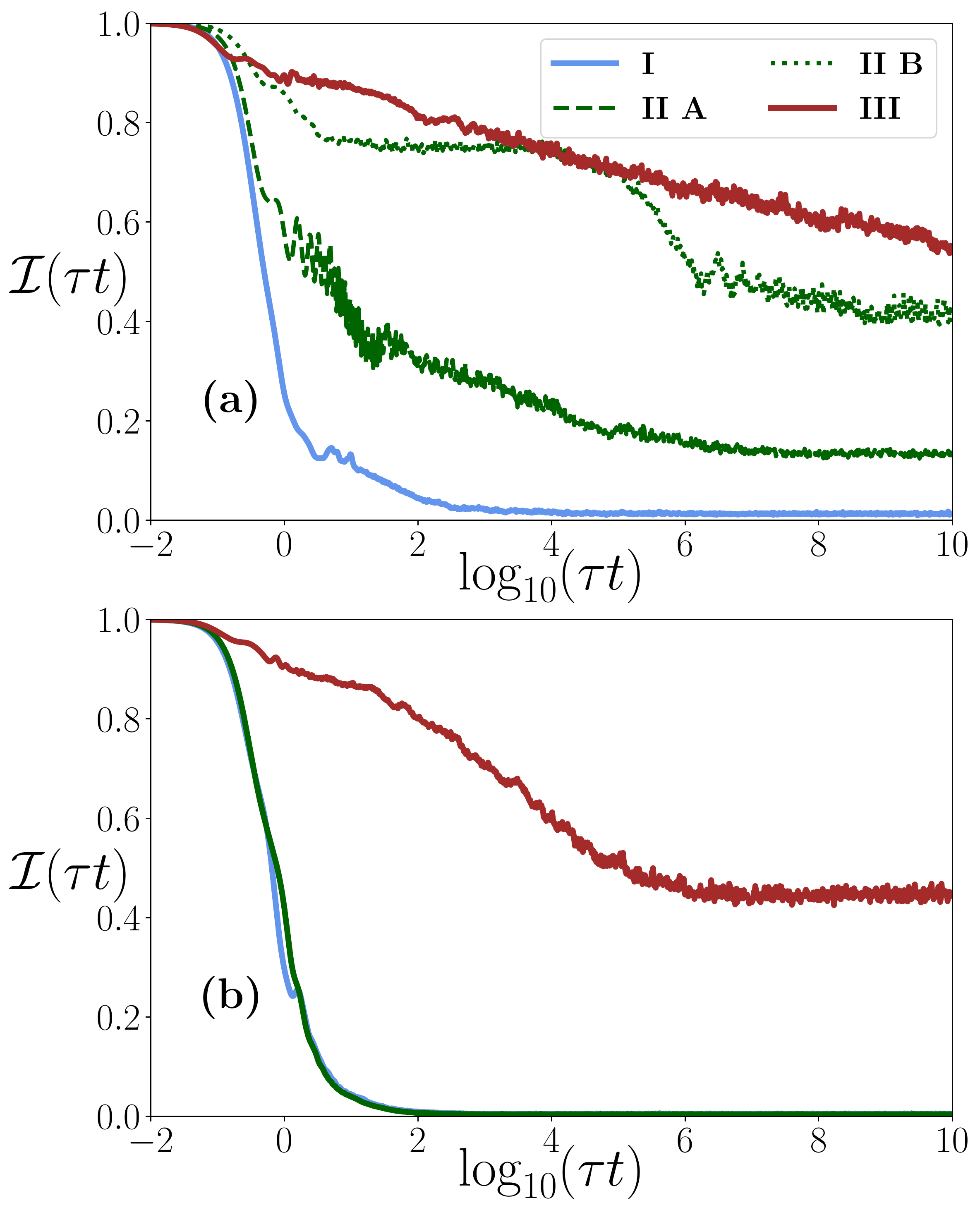}
\caption{(a) Inhomogenity $\mathcal{I}$ as a function of time, $\tau$ for $T/t=-1$ ($V/t=8.8$ for $s=8$), for initial Fock states in the $N_{NN}=L/6$ sector. Family $\textbf{I}$ corresponds to hard-core 
states; family $\textbf{II}$ consists of initial states with a single doubly-occupied states, and family $\textbf{III}$ is formed by the remaining initial Fock states with higher fillings per site. The family $\textbf{II}$ is further split into sub-family {\bf A} , which contains states with a single $\vert \cdots 12 \cdots \rangle$ or $\vert \cdots 21 \cdots \rangle $ arrangement, and sub-family {\bf B} which does not consist of such Fock states. 
Panel (b) shows the results without DDT for the same initial states. In all cases, $L=12$.}
\label{fig:8}
\end{figure} 



\subsection{Time dynamics}

The spectral properties at $T/t=-1$ translate into a markedly different dynamics for the hard-core and the soft-core sectors. Figure~\ref{fig:8}(a) shows (for $L=12$, $s=8$, and $V/t=8.8$) the long-time evolution of the inhomogeneity 
for different initial states within the 
 $N_{NN}=L/6$ sector. The hard-core sector~(family {\bf I}) behaves ergodically, with $\mathcal{I}$ practically vanishing for times $\tau>200/t$. 
 
In the soft-core sector, states with a single double-occupancy~(family {\bf II}) decay much slower. Those with a single pair $\vert \cdots 12 \cdots \rangle$~(sub-family {\bf II A}) present a rapid initial decay of $\mathcal{I}$ corresponding to the spreading of the remaining three bosons over the available space, resembling family {\bf I}. After a short time~(of the order of $1/t$), a slower decay of  $\mathcal{I}$ takes place determined by the highly non-resonant mixing of the occupied pair with the rest. The rest of family {\bf II}, with a single double-occupancy surrounded by empty sites, decays much slower already at short times and then presents a pronounced plateau (at this stage a single occupancy still survives the dynamics) finally reaching a nonzero value. The rest of the soft-core states~(family {\bf III}) is characterized by a single large occupancy, and presents a very slow dynamics.

 In absence of DDT~(Fig.~\ref{fig:8}(b)) initial states belonging to families {\bf I} or {\bf II} undergo a rapid homogenization. Interestingly, only a partial homogenization occurs for a single high-occupancy initial states. This is due to the energy penalty induced by the on-site interaction~\cite{Carleo12}.

Finally, we note that the splitting of the Hilbert space into decoupled sectors results solely from the $T/t=-1$ condition, being independent of the character of interactions. Since it is independent of the considered $1/r^3$ tail, we expect a similar effect for any long-range potential.


\section{Conclusions} 
\label{sec:Conclusions}
Our results show that interaction-induced hopping should play a crucial role in future experiments on the dynamics of polar lattice gases. Density-dependent tunneling strongly modifies the dynamics of soft-core dipolar bosons in one-dimensional lattices. For shallow-enough lattices, the delocalizing effect resulting from the interaction-induced hopping overcomes the localization effect induced 
by the inter-site interactions. As a result, counter-intuitively and in stark contrast to the hard-core case, delocalization is strengthen when the dipolar strength increases. Interestingly, although this is 
generally the case, at a critical dipole strength the density-dependent hopping quasi-cancels the bare hopping resulting in a separation of the Hilbert space in ergodic hard-core states and strongly non-ergodic soft-core ones. 
%


\acknowledgements

M.\L. and J.Z. thank R. Kraus and G. Morigi for discussions on density dependent tunnelings and Wannier function integrals.
We acknowledge support of National Science Centre (Poland) via Opus grants 2019/35/B/ST2/00034 (A.S.A.) and 2019/35/B/ST2/00838 (M.\L{}.). This research was also funded by National Science Centre (Poland) under the OPUS call within the WEAVE programme 2021/43/I/ST3/01142 (J.Z.)
as well as by the
Deutsche Forschungsgemeinschaft (DFG, German Research Foundation) -- Project-ID 274200144 -- SFB 1227 DQ-mat within the project A04, and under Germany's Excellence Strategy -- EXC-2123 Quantum-Frontiers -- 390837967. A partial support by the Strategic Programme Excellence Initiative at Jagiellonian University is also acknowledged. Some of the numerical computations have been possible thanks to PL-Grid Infrastructure.  For the purpose of Open Access, J.Z. has
applied a CC-BY public copyright licence to any Author
Accepted Manuscript (AAM) version arising from this
submission.


\appendix 

\section{Parameters of the extended Bose-Hubbard model}
\label{sec:AppA}

The calculation of the parameters of the extended Bose-Hubbard 
model closely parallels the technique described in detail in Ref.~\cite{Kraus22}. We assume 
a quasi one-dimensional model with an optical lattice along $x$, and a tight harmonic confinement in the transversal directions, leading to the single-particle trapping potential
\begin{equation}
V_t(\mathbf{r}) = \frac{m\omega^2}{2} \left(y^2 + z^2 \right)+ V_0\cos^2(k x)\ ,
\label{pot1}
\end{equation}
where $m$ is the particle mass, $\omega$ is the harmonic trapping frequency along $y$ and $z$, and $k$ is the wavevector of the laser that forms the optical lattice. 

The Hamiltonian of the system may be expressed as~(see e.g. \cite{Dutta15}):
\begin{eqnarray}
\hat{H}&=&\int d^3\mathbf{r} \hat{\Psi}^\dagger (\mathbf{r})\left[-\frac{\hbar^2}{2m}\nabla^2+V_{\text{t}}(\mathbf{r})\right]\hat{\Psi}(\mathbf{r})\\
&+&\frac{1}{2}\int d^3\mathbf{r} \int d^3\mathbf{r}'\hat{\Psi}^\dagger(\mathbf{r})\hat{\Psi}^\dagger (\mathbf{r}')V_{\rm int}(\mathbf{r}-\mathbf{r}')\hat{\Psi}(\mathbf{r}')\hat{\Psi}(\mathbf{r}) \nonumber
\ ,  \label{secquant}
\end{eqnarray}
with  the bosonic field operators $\hat{\Psi}(\mathbf{r})$ and $\hat{\Psi}^\dagger(\mathbf{r})$ that obey the commutation relation $\left[\hat{\Psi}(\mathbf{r}), \hat{\Psi}(\mathbf{r}')^\dagger \right]=\delta ^3\left(\mathbf{r}-\mathbf{r}' \right)$. $V_{\rm int}(\mathbf{r}-\mathbf{r}')$ describes interactions between bosons that is conveniently split into the contact and dipole-dipole terms: 
\setcounter{equation}{3}
\begin{equation}V_{\rm int}(\mathbf{r})=V_c(\mathbf{r})+V_d(\mathbf{r})\,. 
\label{pot2}
\end{equation}
The contact term is characterized by the $s$-wave scattering length, $a_s$. Using the customary notation,   $V_c(\mathbf{r})=g\delta^{(3)}(\mathbf{r})$,  with $g=4\pi \hbar ^2 a_s/m$. Dipole-dipole interactions give a second interaction term,  $V_d(\mathbf{r})$.  We consider dipoles polarized by an external field along the $z$ axis (perpendicular to the axis of the optical lattice) with
\begin{equation}
V_d(\mathbf{r})=C \frac{1-3\cos^2(\theta)}{r^3} \,,
\end{equation}
where $\theta$  is the angle between the dipole and $\mathbf{r}$.
The dipole-dipole interaction is anisotropic in space since the force depends on the dipole orientation.
The  strength of the dipole-dipole interactions $C=\mu_0\mu^2/4\pi$ ($C=d^2/(4\pi\epsilon_0)$) for magnetic (electric) dipoles with moment $\mu$~($d$) where $\mu_0$ ($\epsilon_0$) are the magnetic (electric) permeability, respectively. 

For sufficiently deep optical lattice ($s=V_0/E_R > 3$, where $E_R=\hbar^2k^2/(2m)$) we may expand the field operator as 
\begin{align}\label{exp}
\hat \Psi(\mathbf{r})=\sum_{j=1}^L {\cal W}_j(\mathbf{r}) \hat{a}_{j}  = \sum_{j=1}^L \phi_0(y)\phi_0(z)W_j(x)\, \hat{a}_{j}\,,
\end{align} 
where $j=1,\ldots,L$ denotes the site index, and the operator $\hat{a}_{j}$ annihilates boson at site $j$. The corresponding basis function ${\cal W}_j(\mathbf{r})$ is the product of the ground states of the harmonic oscillators along $y,z$, 
and the Wannier function~(of the lowest band) along the lattice (shallower lattices may implicate the necessity of taking higher bands into account). Plugging Eq.~\eqref{exp} into Eq.~\eqref{secquant}, one expresses the Hamiltonian in as a polynomial of the annihilation and creation operators. Employing the orthogonality of the Wannier functions one arrives at the form:
\begin{align}
\hat{{H}} = -t\sum_{j=1}^{L-1}(\hat{a}^{\dagger}_j\hat{a}_{j+1}+H.c.)
+\frac{1}{2}\sum_{i,j,k,l}^LV_{ijkl}\hat{a}^{\dagger}_i\hat{a}^{\dagger}_j\hat{a}_{k}\hat{a}_{l}.
\label{eqgen}
\end{align}
where the integrals $V_{ijkl}$ are explicitly given as
\begin{align}
V_{ijkl}=\int d^3\mathbf{r} d^3\mathbf{r}'{\cal W}_i(\mathbf{r}) {\cal W}_j(\mathbf{r}')V_{\rm int}(\mathbf{r}-\mathbf{r}'){\cal W}_k(\mathbf{r}'){\cal W}_l(\mathbf{r}).
\label{coe1}
\end{align}
The single particle tunneling amplitude is obtained from the single particle part of the Hamiltonian 
\begin{align}
t=-\int d^3\mathbf{r}{\cal W}_i(\mathbf{r} \left[-\frac{\hbar^2}{2m}\nabla^2+V_t(\mathbf{r})\right]{\cal W}_{i+1}(\mathbf{r})\nonumber \\
= \int dx W_i(x)\left[ \frac{\hbar^2}{2m}\frac{\partial^2}{\partial x^2}- V_0\cos^2(kx)\right]W_{i+1}(x) \ ,
\end{align}
where we limit ourselves to NN tunneling only assuming a sufficiently deep optical lattice
(for shallow lattices, with $s=V_0/E_R<4$, one might need to include next-to-NN tunnelings into the picture, see e.g. 
Ref.~\cite{Trotzky12}.)

As it turns out the integrals over perpendicular directions can be explicitly carried out \cite{Sinha07,Deuretzbacher10,Deuretzbacher13,Bartolo13} yielding
\begin{align}
V_{ijkl}=\int dx dx'W_i(x) W_j(x')V_{\rm 1D}(x-x')W_k(x)W_l(x'),
\label{coev}
 \end{align}
where the effective quasi-one-dimensional potential is given by
\begin{widetext}
\begin{align}
 V_{\rm 1D} = \left(g_{1D}-\frac{2C}{3l^2}\right)\delta(|x-x'|)+ 
 \frac{C}{l^3}\left[\sqrt{\frac{\pi}{8}}e^{(x-x')^2/(2l^2)}\left(1+\frac{(x-x')^2}{l^2}\right){\textrm{Erfc}}\left(\frac{|x-x'|}{l\sqrt{2}}\right)-\frac{|x-x'|}{2l}\right], 
 \label{ugly}
\end{align}
\end{widetext}
in terms of the harmonic oscilator length in the perpendicular direction, $l=(\hbar/m\omega)^{1/2}$
and the effective 1D contact interaction strength $g_{1D}=g/(2\pi l^2)$. In Eq.~\eqref{ugly} $\delta$ stands for Dirac delta function and Erfc for the complementary error function. Note that the term proportional to the delta function contains contributions from both the contact and dipole-dipole interactions.

The largest contribution is given by the on-site 
interaction term~(diagonal in $ijkl$ indices), traditionally denoted as $U\equiv V_{iiii}$. For contact interactions this is the dominant term. For long-range dipolar interactions the next important term has the form of a density-density
interaction $(V_{ijij}+V_{ijji}) {\hat n}_i{\hat n_j}$ (for $i\ne j$) where often only the NN term for $ j=i\pm 1$ is taken into account. 
We mention parenthetically that while for contact interactions  $V_{ijij}$ and $V_{ijji}$ are identical, for a dipolar potential one finds that $|V_{ijij}|\ll |V_{ijji}|$. Since Wannier functions are well localized on sites, for tight perpendicular binding  and $1/r^3$ potential one may approximate $V_{ijij}=V/|i-j|^3$ recovering the 
typical dipolar tail; here {$V= V_{0110}$} is the value of the integral for the NNs. A standard extended Bose-Hubbard model (see e.g. \cite{Rossini12}) considers just terms involving $U$ and 
$V$ coefficients and neglects the dipolar tail. The latter may play an important role in the dynamics of the system \cite{Li21,Korbmacher23}, and may differ from the 
standard $1/r^3$ decay if the transversal confinement 
is not sufficiently strong~\cite{Korbmacher23}.

Other important terms, introduced by Hirsch \cite{Hirsch94} for strongly-correlated spinful fermions, are density-dependent tunnelings~(DDT), also called correlated hoppings, coming from $V_{ijkl}$ terms with three equal indices. The most important corresponds to NN correlated tunneling, e.g. $V_{iii(i+1)} {\hat a}_i^\dagger {\hat n}_i {\hat a}_{i+1}$. For shortness of notation the corresponding amplitude is denoted by $T$ (or rather, due to some historical reasons $-T$ \cite{Dutta15}). Taking together contributions containing $U$, $V$, and $T$ terms, one arrives at the extended Bose-Hubbard Hamiltonian of Eq.~\eqref{eq:ext}.

%

\end{document}